\documentclass{iau}
\usepackage{graphicx}
\usepackage{subcaption}
\title[Mean Magnetic Field (MMF)] 
{Role of the background regimes towards the Solar Mean Magnetic Field}

\author[Bose \& Nagaraju]   
{Souvik Bose$^1$$^,$$^2$
 \and K. Nagaraju$^3$}

\affiliation{$^1$Institute of Theoretical Astrophysics, University of Oslo, P.O. Box 1029 Blindern, N-0315 Oslo, Norway \\ $^2$Rosseland Centre for Solar Physics, University of Oslo, P.O. Box 1029 Blindern, N-0315 Oslo, Norway email: {\tt souvik.bose@astro.uio.no} \\[\affilskip]
$^3$Indian Institute of Astrophysics, Koramangala-2nd Block, Bangalore-560034 }

\pubyear{2018}
\volume{340}  
\pagerange{119--126}
\jname{Long Term Datasets for the Understanding of Solar and Stellar
Magnetic Cycle}
\editors{Dipankar Banerjee, Jie Jiang, Kanya Kusano \& Sami Solanki}
\begin{document}

\maketitle

\begin{abstract}
The Solar Mean Magnetic Field (SMMF) is generally defined as disc-averaged line-of-sight (LOS) magnetic field on the sun. The role of active regions and the large-scale magnetic field structures (also called the background) has been debated over past few decades to understand whether the origin of SMMF is either due to the active regions or the background. In this paper, we have investigated the contribution of sunspots, plages, network regions and the background towards the SMMF using data from the SDO-AIA \& HMI, and found that 83\% of the SMMF is due to the background whereas the remaining 17\% originates from the active and network regions.
\end{abstract}

\firstsection 
\section{Introduction}
Studying the SMMF gives an idea of \textit{sun-as-a-star} magnetic field and its variation. It also reflects
the imbalance in magnetic flux
of opposite polarity on the visible disk
(\cite[Svalgaard \etal\ (1975)]{Svalgaard_etal75}). Several studies like \cite[Scherrer \etal\ (1977)]{Scherrer_1977} \& \cite[Gough \etal\ (2017)]{Gough_2017} have suggested that the imbalance is primarily due to the large-scale magnetic field structure, also termed as the background magnetic field whereas recent investigations led by \cite[Kutsenko et.al.(2017)]{Kutsenko_2017} suggests that the active regions (sunspots) play a pivotal role in the SMMF. In this paper, we followed a conventional approach of detecting the sunspots, plages, and the network features using intensity images from SDO-AIA 1600 \AA\, and 4500 \AA\, datasets and thereby studied the contribution towards the SMMF with HMI.720s LOS magnetograms.

\section{Contribution of various surface magnetic features to the observed SMMF}

\begin{figure}[htb]
\centering
  \begin{tabular}{@{}cc@{}}
  \includegraphics[width=8.5cm]{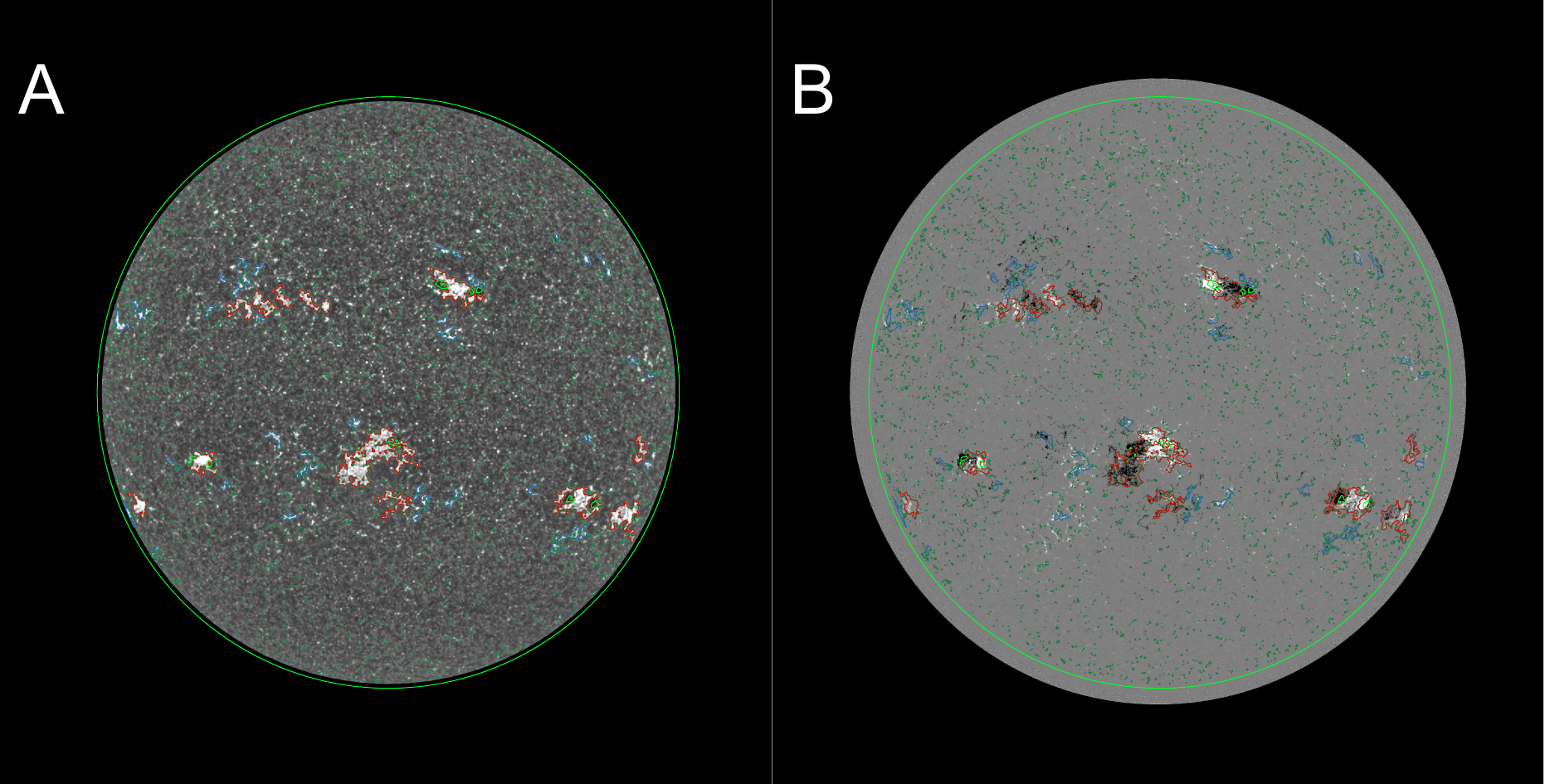}&
  \includegraphics[width=4.5cm,angle=90]{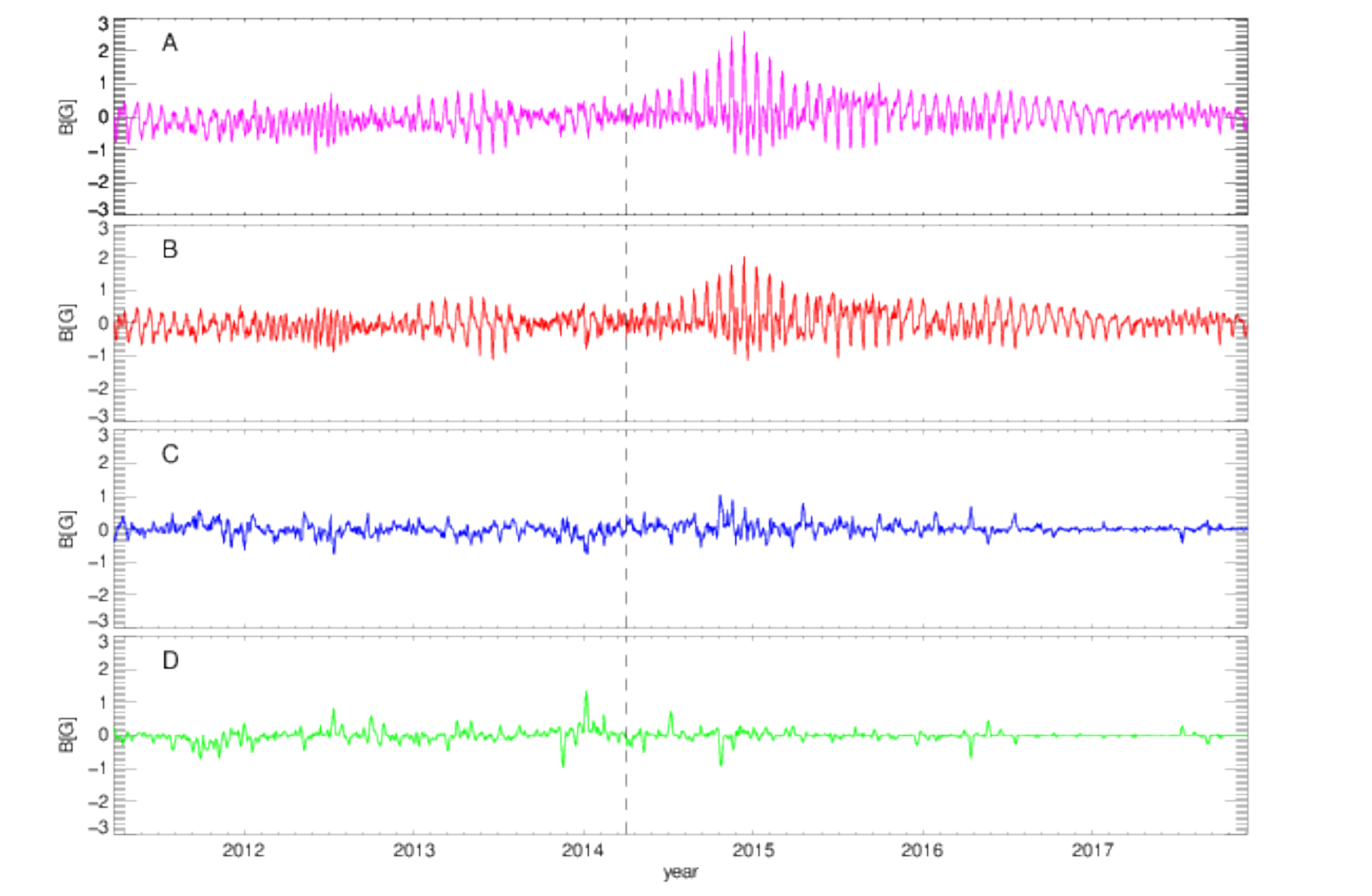}
  \end{tabular}
  \caption{Left panel:Automated detection of various features on:(A) AIA 1600 \AA\, data \& (B) Corresponding HMI Magnetogram. Right Panel: Magnetic field corresponding to the various segments-(A) SMMF, (B) background field, (C) plage and network fields, \& (D) sunspot field.}
  \label{Fig1}
\end{figure}

We developed an automated detection algorithm that segregated plage \& enhanced network regions from AIA 1600 \AA\, images following an adaptive intensity thresholding technique based on $\mu_{image}$ + K  $\sigma_{image}$ (where $\mu$ is the mean, $\sigma$ is the standard deviation and $K$ = $1.71$) and imposing an area threshold criterion. Further, we segregated the active networks and sunspots from AIA 1600 \AA\, \& 4500 \AA\, respectively, by implementing only the adaptive thresholding procedure. A sample detection is shown in the left panel of figure \ref{Fig1}. The details of the algorithm and the processing steps can be found in \cite[Bose \& Nagaraju (2018)]{Bose_2018}. 

The pixels corresponding to each of the regions were grouped into three categories, namely (1) sunspots, (2) plage, enhanced and active network regions (as one entity) and (3) background regions that do not belong to either (1) or (2). The magnetic field contributions corresponding to each of these regions towards the SMMF was computed using linear regression analysis (\cite[Bose \& Nagaraju (2018)]{Bose_2018}) with the HMI LOS magnetograms between 21.03.2011 \& 30.11.2017.

The right panel of figure \ref{Fig1}-(A), shows the temporal variation of the SMMF. It is found to have a peak value of  about 2.5~G that is consistent with \cite[Kutsenko et.al.(2016)]{Kutsenko_2016}. The background field (figure \ref{Fig1}-(B)) emulates the SMMF quite distinctly, both peaking around December 2014. However, it is clear that the fields corresponding to the plage, network regions and the sunspots, as in figure \ref{Fig1}-(C) \& (D), has no correlation with the SMMF whatsoever. This is in contrast with \cite[Kutsenko et.al.(2017)]{Kutsenko_2017}, however it is in agreement with the earlier pioneering works of \cite[Svalgaard \etal\ (1975)]{Svalgaard_etal75} \& \cite[Scherrer \etal\ (1977)]{Scherrer_1977}.

A linear regression analysis shows that the total plages and the network fields contribute only about 14\% towards the SMMF with a Pearson R of $0.316$, whereas the background contributes about 83\% with Pearson R of $0.943$. The contribution of the sunspots is found to be statistically insignificant at 95\% confidence level. Details can be found in \cite[Bose \& Nagaraju (2018)]{Bose_2018}.

\section{Conclusions}

The variation of the line-of-sight field corresponding to the different segments clearly suggest that the background field is the major contributor to the SMMF while the active regions including the network fields provide a minimal contribution. Based on these results, we conclude that the origin of the SMMF lies in the large-scale magnetic field structures on the sun.


\begin{thebibliography}{}

 \bibitem[Svalgaard \etal\ (1975)]{Svalgaard_etal75}
{L.~Svalgaard, J.~M.~Wilcox and P.~H.~Scherrer} 1975
{\em Solar Physics}, 45, 83

\bibitem[Gough \etal\ (2017)]{Gough_2017}
{D.~O. Gough.} 2017
{\em Solar Physics}, 292,70

\bibitem[Scherrer \etal\ (1977)]{Scherrer_1977}
{P.~H.~Scherrer, J.~M.~Wilcox, V.~Kotov, A.~B.~Severny \& R.~Howard} 1977,
{\em Solar Physics}, 52,6

\bibitem[Kutsenko \& Abramenko (2017)]{Kutsenko_2017}
{A.~S.~Kutsenko, V.~I.~Abramenko \& V.~B.~Yurchyshyn} 2017,
{\em Solar Physics},292,121

\bibitem[Bose \& Nagaraju (2018-submitted to ApJ)]{Bose_2018}
 {Souvik Bose \& K. Nagaraju} 2018
 {\em Submitted to the Astrophysical Journal} 

\bibitem[Kutsenko \& Abramenko (2016)]{Kutsenko_2016}
{A.~S.~Kutsenko \& V.~I.~Abramenko} 2016,
{\em Solar Physics},291,1613


\end{thebibliography}
\end{document}